\documentclass[preprint,showpacs,amsmath,amssymb,eqsecnum]{revtex4}
\usepackage{dcolumn}
\usepackage{bm}
\usepackage{graphicx}

\makeatletter

\def\compoundrel#1\over#2{\mathpalette\compoundreL{{#1}\over{#2}}}
\def\compoundreL#1#2{\compoundREL#1#2}
\def\compoundREL#1#2\over#3{\mathrel
  {\vcenter{\hbox{$\m@th\buildrel{#1#2}\over{#1#3}$}}}}
\makeatother
\newcommand{\be}{\begin{equation}}
\newcommand{\ee}{\end{equation}}
\newcommand{\bea}{\begin{eqnarray}}
\newcommand{\eea}{\end{eqnarray}}

\newcommand{\mapright}[1]{%
	\smash{\mathop{%
	\hbox to 1cm{\rightarrowfill}}\limits^{#1}}}

\begin{document}
\draft
\title{{Prediction for Quark Mixing from Universal Quark and Lepton Mass Matrices with Flavor 2 \(\leftrightarrow\) 3 Symmetry}}

\author{Koichi MATSUDA}
\affiliation{%
Graduate school of Science, 
Osaka University, Toyonaka, Osaka 560-0043, Japan}
\author{Hiroyuki NISHIURA}
\affiliation{%
Department of General Education, 
Junior College of Osaka Institute of Technology, \\
Asahi-ku, Osaka 535-8585, Japan}

\date{Jan 19, 2005}

\begin{abstract}

We propose a mass matrix model that gives a unified description of 
quark and lepton with the same texture form based on a flavor $2 \leftrightarrow 3$ symmetry.
The model is in contrast with the conventional picture that the mass matrix forms in the quark sector will 
take somewhat different structures from those in the lepton sector.
By investigating possible types of assignment for masses, 
we find that our quark mass matrices can 
lead to small CKM quark mixings which are consistent with experimental values. 
We also find some phase-parameter-independent relations among 
CKM quark mixing matrix elements. 
\end{abstract}
\pacs{12.15.Ff, 11.30.Hv, 11.10.Hi}


\maketitle

\section{Introduction}

Recent neutrino oscillation experiments~\cite{skamioka} have highly
suggested a nearly bimaximal lepton mixing $(\sin^2 2\theta_{12}\sim 1$, 
$\sin^2 2\theta_{23}\simeq 1)$ in contrast with the small quark mixing.
In order to reproduce these large lepton mixing and small quark mixing, 
mass matrices of various structures with texture zeros have been investigated 
in the literature~\cite{fritzsch}-\cite{Ramond}.
Recently a following mass matrix model based on a discrete symmetry $Z_3$ and 
a flavor $2 \leftrightarrow 3$ symmetry has been proposed~\cite{Koide} 
for all quarks and leptons:  
\begin{equation}
M=P^\dagger \left(
\begin{array}{lll}
\ 0 & \ A & \ A \\
\ A & \ B & \ C \\
\ A & \ C & \ B \\
\end{array}
\right) P^\dagger, \ \ 
\end{equation}
where $A$, $B$, and $C$ are real parameters. 
The diagonal phase matrix $P$ which breaks $2 \leftrightarrow 3$ symmetry has been introduced from a phenomelogical points of view.
This structure of mass matrix has been previously suggested and used 
for the neutrino mass matrix in Refs.~\cite{Fukuyama}--\cite{Nishiura} 
motivated by the experimental finding of maximal \(\nu_\mu\)--\(\nu_\tau\) 
mixing~\cite{skamioka}.  We consider that this structure 
is fundamental for both quarks and leptons, although it was speculated 
from the neutrino sector. Therefore, we assume that all the mass matrices 
have this structure, which is against the conventional picture that the mass matrix forms 
in the quark sector will 
take somewhat different structures from those in the lepton sector.
In our previous works \cite{Koide,Matsuda,Matsuda2}, 
we have pointed out that this structure leads to reasonable values for 
the Cabibbo--Kobayashi--Maskawa (CKM) \cite{CKM} quark mixing 
as well as large lepton mixing. 
However, the diagonal phase matrix $P$ 
has been introduced artificially into the model. 
\par 
Recently, a new way to include phases into the mass matrix $M$ has been proposed by Koide~\cite{Koide2} as 
\begin{equation}
M  =\left(
\begin{array}{lll}
\ 0 & \ ae^{-i\phi} & \ a \\
\ ae^{-i\phi} & \ be^{-2i\phi} & \ (1-\xi)b \\
\ a & \ (1-\xi)b & \ b \\
\end{array}
\right) , 
\end{equation}
where $a$, $b$, and $\xi$ are real parameters and $\phi$ is a phase. 
$M$ is invarient under the "extended " flavor $2 \leftrightarrow 3$ permutation $T^{\phi}_{23}$ for the fields: 
\begin{equation}
T^{\phi}_{23} M \left(T^{\phi}_{23}\right)^T=M , 
\end{equation}
with
\begin{equation}
T^{\phi}_{23} =\left(
\begin{array}{lll}
\ 1 \ & \ 0 & \ 0 \\
\ 0 \ & \ 0 & \ e^{-i\phi} \\
\ 0 \ & \ e^{i\phi} & \ 0 \\
\end{array}
\right) . 
\end{equation}
Koide~\cite{Koide2} has discussed the democratic case ($\xi=0$ ) and obtained CKM mixing matrix which is roughly in agreement with experimental values.
However, $|V_{cb}|$ is somewhat smaller than the observed value.
\par
In the present paper, extending the Koide's model, we propose a nonsymmetric 
and nondemocratic mass matrix by adding a new parameter $a^\prime$ to 
reproduce a consistent value for $|V_{cb}|$.   
Our mass matrix form is  
\begin{equation}
M  =\left(
\begin{array}{lll}
\ 0 & \ ae^{-i\phi} & \ a \\
\ a^\prime e^{-i\phi} & \ be^{-2i\phi} & \ (1-\xi)b \\
\ a^\prime & \ (1-\xi)b & \ b \\
\end{array}
\right) . \label{ourM}
\end{equation}
In order to avoid difficulties raised in Ref~\cite{Koide3} in considering flavor symmetry, 
we consider that the $\phi$ is a phase which breaks the 2 \(\leftrightarrow\) 3 symmetry, 
although it is originally introduced by the "extended"  2 \(\leftrightarrow\) 3 symmetry. 
In this paper, we shall investigate CKM quark mixing from the phenomenological point of view 
by using the mass matrix given in Eq.~(\ref{ourM}).  
\par
This article is organized as follows. 
In Sec.~II, we discuss the mass matrix of our model. 
The numerical values of CKM quark mixing matrix elements at the unification scale are obtained from the observed values at the electroweak scale in Sec.~III. 
The quark mixing matrix in the present model is argued in Sec.~IV.  
Sec.~V is devoted to a summary.

\section{Mass matrix}
\par 
Our mass matrices 
\(M_u\), \(M_d\), \(M_\nu\), and  \(M_e\) 
for up quarks (\(u,c,t\)), down quarks (\(d,s,b\)), 
neutrinos (\(\nu_e,\nu_\mu,\nu_\tau\)) and 
charged leptons (\(e,\mu,\tau\)), 
respectively are given as follows:
\begin{equation}
M_f  =\left(
\begin{array}{lll}
\ 0 & \ a_fe^{-i\phi_f} & \ a_f \\
\ a^\prime_fe^{-i\phi_f} & \ b_fe^{-2i\phi_f} & \ (1-\xi_f)b_f \\
\ a^\prime_f & \ (1-\xi_f)b_f & \ b_f \\
\end{array}
\right) ,\quad \mbox{($f=u,d,\nu$, and $e$)} \label{ourMf}
\end{equation}
where $a_f$, $b_f$, $\xi_f$, and $a^\prime_f$ are real parameters and $\phi_f$ is a phase parameter. 
Here we consider a nonsymmetric mass matrix based on the flavor 
2 \(\leftrightarrow\) 3 symmetry with including one symmetry-breaking phase parameter. 
From Eq.~(\ref{ourMf}), hermitian matrix $M_f M_f^\dagger$ can be decomposed to 
following two parts:
\begin{equation}
M_f M_f^\dagger
= 
2a_f^2\left(
\begin{array}{lll}
\ 1 & \ 0 & \ 0 \\
\ 0 & \ 1 & \ 0 \\
\ 0 & \ 0 & \ 1 \\
\end{array}
\right) 
+
P_{f}^\dagger \left(
\begin{array}{lll}
\ 0 & \ A_f & \ A_f \\
\ A_f & \ B_f & \ C_f \\
\ A_f & \ C_f & \ B_f \\
\end{array}
\right) 
P_{f}, \ \ \label{MM}
\end{equation}
where real components $A_f$, $B_f$, and $C_f$ are given by 
\begin{eqnarray}
A_f & =&a_f|X_f|  ,\label{eq1001}\\
B_f & =&(a_f^\prime)^2-2a_f^2+b_f^2+(1-\xi_f)^2b_f^2 ,\label{eq1002}\\ 
C_f & =&(a_f^\prime)^2+(1-\xi_f)b_f^2(e^{i\phi_f}+e^{-i\phi_f}) \label{eq1003}, 
\end{eqnarray}
and the diagonal phase matrix $P_{f}$ is given by
\begin{equation}
P_{f} =\mbox{diag}\left(1,e^{i(\phi_f-\varphi_f)},e^{-i\varphi_f}\right) \label{pf}.
\end{equation}
Here we have defined $|X_f|$ and phase $\varphi_f$  as 
the absolute and phase parts of $X_f \equiv b_f+(1-\xi_f)b_fe^{i\phi_f}$, respectively: 
\begin{equation}
X_f \equiv b_f+(1-\xi_f)b_fe^{i\phi_f}\equiv|X_f|e^{i\varphi_f}. \ \label{Xdef} 
\end{equation}
\par
First we discuss three mass eigenvalues $m_{if}$ (i=1,2, and 3) of $M_f$. 
From Eq.~(\ref{MM}), the eigenvalues $m_{if}^2$ of $M_f M_f^\dagger$ are obtained in terms of the components as 
\begin{eqnarray}
m_{1f}^2 & =&\frac{1}{2}
\left[B_f+C_f-\sqrt{8A_f^2 + (B_f+C_f)^2}
\right]+2a_f^2, \\
m_{2f}^2 & =&\frac{1}{2}
\left[B_f+C_f+\sqrt{8A_f^2 + (B_f+C_f)^2}
\right]+2a_f^2,\\
m_{3f}^2 & =&(B_f-C_f)+2a_f^2,
\end{eqnarray}  
from which, by using Eqs.(\ref{eq1001})-(\ref{eq1003}), we obtain 
\begin{eqnarray}
m_{1f} & =& \frac{1}{2}\left[\sqrt{|X_f|^2+2(a_f^\prime-a_f)^2}-\sqrt{|X_f|^2+2(a_f^\prime+a_f)^2}\right]<0 ,\label{eq1201} \\ 
m_{2f} & =& \frac{1}{2}\left[\sqrt{|X_f|^2+2(a_f^\prime-a_f)^2}+\sqrt{|X_f|^2+2(a_f^\prime+a_f)^2}\right]>0 , \label{eq1202}\\ 
m_{3f} & =& \sqrt{b_f^2+(1-\xi_f)^2b_f^2-2(1-\xi_f)b_f^2 \cos\phi_f}>0. \label{m3}
\end{eqnarray}
There are five parameters, $a_f$, $a^\prime_f$, $b_f$, $\xi_f$, and $\phi_f$ in $M_f$. 
If we fix above three eigenvalues $m_{if}$ by the quark masses, 
we have only two free parameters.  
As the free parameters, let us chose a parameter $\alpha_f$ defined by
\begin{eqnarray}
\alpha_f\equiv\frac{a_f^\prime}{a_f} ,\label{defalpha}
\end{eqnarray}
and a phase parameter $\eta_f$ defined in Fig.~1 independently of mass eigenvalues \(m_{if}\). 
Then, from Eqs.~(\ref{eq1201}), (\ref{eq1202}), and (\ref{defalpha}), 
the parameters $a_f$ and $|X_f|$ are expressed in term of $\alpha_f$ as
\begin{eqnarray}
a_f^2& =&\frac{|m_{1f}|m_{2f}}{2\alpha_f},\label{af^2-A}\\
|X_f|^2 & =&m_{1f}^2+m_{2f}^2
			-|m_{1f}|m_{2f}\left(\frac{1+\alpha_f^2}{\alpha_f}\right) \nonumber\\
& =&-\frac{|m_{1f}|m_{2f}}{\alpha_f}\left(\alpha_f - \frac{|m_{1f}|}{m_{2f}}\right)
	 \left(\alpha_f-\frac{m_{2f}}{|m_{1f}|}\right). \label{Xf^2-A} 
\end{eqnarray}
From Fig.~1 with Eqs.~(\ref{Xdef}) and (\ref{m3}), the parameters  
$\varphi_f$, $\phi_f$, $b_f$, and $(1-\xi_f)b_f$ are expressed by $|X_f|$ and $\eta_f$ as 
\begin{eqnarray}
\cos\varphi_f & =&\frac{|X_f|-m_{3f}\cos\eta_f}
			{\sqrt{|X_f|^2+m_{3f}^2-2m_{3f}|X_f|\cos\eta_f}} ,\label{eq3001}\\
\cos\left(\phi_f-\varphi_f\right) & =&\frac{|X_f|+m_{3f}\cos\eta_f}
		{\sqrt{|X_f|^2+m_{3f}^2+2m_{3f}|X_f|\cos\eta_f}} ,\label{eq3002}\\
\cos\phi_f & =&\frac{|X_f|^2-m_{3f}^2}
	{\sqrt{\left(|X_f|^2+m_{3f}^2\right)^2-4m_{3f}^2|X_f|^2\cos^2 \eta_f}} ,\label{eq3003}\\
b_f^2 & =& \frac{1}{4}\left(|X_f|^2+m_{3f}^2-2|X_f|m_{3f}\cos\eta_f\right), \\
(1-\xi_f)^2b_f^2 & =& \frac{1}{4}\left(|X_f|^2+m_{3f}^2+2|X_f|m_{3f}\cos\eta_f\right).
\label{eq010701}
\end{eqnarray}
Therefore, by using Eqs.~(\ref{af^2-A}) - (\ref{eq010701}), 
the all texture components in $M_f$ 
are given by $\alpha_f$ and $\eta_f$. 
It should be noted that from Eq.~(\ref{Xf^2-A})  
we have the bound for the parameter $\alpha_f$ as 
\begin{equation}
\frac{|m_{1f}|}{m_{2f}}\le \alpha_f \le \frac{m_{2f}}{|m_{1f}|} ,
\end{equation}
and also $|X_f|$ has an upper bound as 
\begin{equation}
0 \le |X_f| \le m_{2f}-|m_{1f}| .
\end{equation} 
\par
Secondly we discuss an unitary matrix $U_{Lf}$ which diagonalizes $M_f M_f^\dagger$.
The explicit expression of $U_{Lf}$ depends on the following 
three types of assignment for $m_i$ as shown in the Appendix:\\ 
\par
(i) Type A:
\par
In this type, the mass eigenvalues \(|m_{1f}|\), \(m_{2f}\), and \(m_{3f}\) 
are allocated to the masses of the first, second, and third generations, respectively.
(i.e. \(|m_{1f}|\)\(\ll\)\(m_{2f}\)\(\ll\)\(m_{3f}\).)
In this type, the $M_f M_f^\dagger$ in Eq.~(\ref{MM}) is diagonalized by an unitary matrix $U_{Lf}$ as
\begin{equation}
U_{Lf}^\dagger M_f M_f^\dagger U_{Lf} = \mbox{diag}\left(m_{1f}^2, m_{2f}^2, m_{3f}^2\right),
\end{equation}
where
\begin{equation}
U_{Lf}=P_{f}^\dagger O_f .
\end{equation}
Here $P_{f}$ is the diagonal phase matrix given in Eq.~(\ref{pf}) and 
$O_f$ is the orthogonal matrix which diagonalizes the real symmetric matrix in the second term in Eq.~(\ref{MM}):
\begin{equation}
O_f^T \left(
\begin{array}{lll}
\ 0 & \ A_f & \ A_f \\
\ A_f & \ B_f & \ C_f \\
\ A_f & \ C_f & \ B_f \\
\end{array}
\right)
O_f=\mbox{diag}(-\lambda_{1f},\lambda_{2f},\lambda_{3f}) .
\end{equation}
Here, the eigenvalues $\lambda_{1f}$, $\lambda_{2f}$, and $\lambda_{3f}$ are given by
\begin{eqnarray}
-\lambda_{1f} & =&m_{1f}^2 -2a_f^2=m_{1f}^2 -\frac{|m_{1f}|m_{2f}}{\alpha_f}, \\
\lambda_{2f} & =& m_{2f}^2 -2a_f^2=m_{2f}^2 -\frac{|m_{1f}|m_{2f}}{\alpha_f}, \\
\lambda_{3f} & =& m_{3f}^2 -2a_f^2=m_{3f}^2 -\frac{|m_{1f}|m_{2f}}{\alpha_f}, 
\end{eqnarray}
and the orthogonal matrix \(O_f\) is expressed as
\begin{equation}
O_f\equiv
\left(
\begin{array}{ccc}
{  c_f}  & {  s_f}& {0} \\
{- \frac{s_f}{\sqrt{2}}} & {\frac{c_f}{\sqrt{2}}} & {-\frac{1}{\sqrt{2}}} \\
{- \frac{s_f}{\sqrt{2}}} & {\frac{c_f}{\sqrt{2}}} & {\frac{1}{\sqrt{2}}}
\end{array}
\right), \label{eq990114} 
\end{equation}
where 
\begin{eqnarray}
c_f & =&\sqrt{\frac{\lambda_{2f}}{\lambda_{2f}+\lambda_{1f}}}
= \sqrt{\frac{m_{2f}^2 -\frac{|m_{1f}|m_{2f}}{\alpha_f}}{m_{2f}^2-m_{1f}^2}}\ ,\\
s_f & =&\sqrt{\frac{\lambda_{1f}}{\lambda_{2f}+\lambda_{1f}}}
=\sqrt{\frac{\frac{|m_{1f}|m_{2f}}{\alpha_f}-m_{1f}^2}{m_{2f}^2-m_{1f}^2}}.
\end{eqnarray}
It should be noted that the mixing angles are functions of only $\alpha_f$, 
since the $m_{if}$ is fixed by the experimental quark mass values. 
We find from Eq.~(\ref{eq3003}) that $\phi_f \approx \pm \pi$ for 
$m_{1f}^2 \ll m_{2f}^2 \ll m_{3f}^2$ in this type A assignment. \\
\par
(ii)\ Type B: 
\par
In this type, the mass eigenvalues \(|m_{1f}|\), \(m_{3f}\), and \(m_{2f}\) 
are allocated to the masses of the first, second, and third generations, respectively.
(i.e. \(|m_{1f}|\)\(\ll\)\(m_{3f}\)\(\ll\)\(m_{2f}\).)
The real symmetric matrix in the second term in Eq.~(\ref{MM}) is 
diagonalized by an orthogonal matrix $O_f^{\prime}$ as follows: 
\begin{equation}
O_f^{\prime T} \left(
\begin{array}{lll}
\ 0 & \ A_f & \ A_f \\
\ A_f & \ B_f & \ C_f \\
\ A_f & \ C_f & \ B_f \\
\end{array}
\right)
O_f^{\prime}=\mbox{diag}(-\lambda_{1f},\lambda_{3f},\lambda_{2f}) .
\end{equation}
Here $O_f^{\prime}$ is obtained from \(O_f\) by exchanging the second row  
for the third one as  
\begin{equation}
O_f^\prime\equiv
\left(
\begin{array}{ccc}
{ c_f} & {0} & { s_f} \\
{- \frac{s_f}{\sqrt{2}}} & {\frac{1}{\sqrt{2}}} & {\frac{c_f}{\sqrt{2}}} \\
{- \frac{s_f}{\sqrt{2}}} & {-\frac{1}{\sqrt{2}}} & {\frac{c_f}{\sqrt{2}}}
\end{array}
\right). \label{eq990114B} 
\end{equation}
Therefore the $M_f M_f^\dagger$ in Eq.~(\ref{MM}) is diagonalized by an unitary matrix $U_{Lf}$ as
\begin{equation}
U_{Lf}^\dagger M_f M_f^\dagger U_{Lf} = \mbox{diag}\left(m_{1f}^2, m_{3f}^2, m_{2f}^2\right),
\end{equation}
where
\begin{equation}
U_{Lf}=P_{f}^\dagger O_f^{\prime} .
\end{equation}
In this type B assignment,  we find $\phi_f\sim 0$ which is in contrast with the case in type A.\\
\par
(iii)\ Type C: 
\par
In this type, the mass eigenvalues \(m_{3f}\), \(|m_{1f}|\), and \(m_{2f}\) 
are allocated to the masses of the first, second, and third generations, respectively.
(i.e. \(m_{3f}\)\(\ll\)\(|m_{1f}|\)\(\ll\)\(m_{2f}\).) In this type, we have 
\begin{equation}
O_f^{\prime \prime T} \left(
\begin{array}{lll}
\ 0 & \ A_f & \ A_f \\
\ A_f & \ B_f & \ C_f \\
\ A_f & \ C_f & \ B_f \\
\end{array}
\right)
O_f^{\prime \prime}=\mbox{diag}(\lambda_{3f},-\lambda_{1f},\lambda_{2f}) ,
\end{equation}
\begin{equation}
U_{Lf}^\dagger M_f M_f^\dagger U_{Lf} = \mbox{diag}\left(m_{3f}^2, m_{1f}^2, m_{2f}^2\right),
\end{equation}
where
\begin{equation}
U_{Lf}=P_{f}^\dagger O_f^{\prime \prime} .
\end{equation}
Here, the orthogonal matrix \(O_f''\) is given by  
\begin{equation}
O_f''\equiv
\left(
\begin{array}{ccc}
{0}                   & { c_f}                   &  { s_f} \\
{\frac{1}{\sqrt{2}}}  & {- \frac{s_f}{\sqrt{2}}} &  {\frac{c_f}{\sqrt{2}}} \\
{-\frac{1}{\sqrt{2}}} & {- \frac{s_f}{\sqrt{2}}} &  {\frac{c_f}{\sqrt{2}}}
\end{array}
\right). \label{eq990114C} 
\end{equation}
This type is not so useful to get the reasonable CKM quark mixing values.

\section{Estimation of the mixing angles at \(\mu=M_X\)}
\par
The mass matrix in Eq. (\ref{ourMf}) holds only at the unification scale \(\mu=M_X\). 
Therefore, in order to compare our model and the experimental data, 
we calculate numerical values of the CKM mixing matrix elements at  \(\mu=M_X\) 
from their observed values at the electroweak scale \(\mu=m_Z\). 
We estimate the evolution effects for the CKM matrix elements 
from \(\mu=m_Z\) to \(\mu=M_X\)
by using the two-loop renormalization group equation (RGE). 
In the numerical calculations, we use the running quark masses which is estimated 
in Ref.~\cite{Fusaoka}
[minimal supersymmetric standard model with tan$\beta$=10 case] : 
\begin{equation}
\begin{array}{lll}
m_u(m_Z)=2.33^{+0.42}_{-0.45}\, \mbox{MeV},& 
m_c(m_Z)=677^{+56}_{-61}\, \mbox{MeV},&
m_t(m_Z)=181\pm13\, \mbox{GeV},\\
m_d(m_Z)=4.69^{+0.60}_{-0.66}\, \mbox{MeV},& 
m_s(m_Z)=93.4^{+11.8}_{-13.0}\, \mbox{MeV},&
m_b(m_Z)=3.00\pm0.11\, \mbox{GeV}.\\ 
\end{array}
\label{eq123103}
\end{equation}
\begin{equation}
\begin{array}{lll}
m_u(M_X)=1.04^{+0.19}_{-0.20}\, \mbox{MeV},& 
m_c(M_X)=302^{+25}_{-27}\, \mbox{MeV},&
m_t(M_X)=129^{+196}_{-40}\,  \mbox{GeV},\\
m_d(M_X)=1.33^{+0.17}_{-0.19}\, \mbox{MeV},& 
m_s(M_X)=26.5^{+3.3}_{-3.7}\, \mbox{MeV},&
m_b(M_X)=1.00\pm0.04\, \mbox{GeV}.\\ 
\end{array}
\label{eq123104}
\end{equation}
The observed quark mixing angles and the $CP$ violating phase at \(\mu=m_Z\) 
are given by \cite{PDG}
\begin{eqnarray}
\sin\theta_{12}(m_Z) &=& 0.2243 \pm 0.0016, \quad \sin\theta_{23}(m_Z) = 0.0413 \pm 0.0015, \nonumber\\
\sin\theta_{13}(m_Z) &=& 0.0037 \pm 0.0005, \quad \delta(m_Z) = 60^\circ \pm 14^\circ. \label{CKM_mz}
\end{eqnarray}
Using the values in Eq.~(\ref{CKM_mz}) as inputs, we obtain the following numerical values at \(\mu=M_X\):
\begin{align}
\sin\theta_{12}^0 &= 0.2226 - 0.2259, \quad 
\sin\theta_{23}^0 = 0.0295 - 0.0383, \nonumber \\
\sin\theta_{13}^0 &= 0.0024 - 0.0038, \quad 
\delta^0    \  = \ 46^\circ   - 74^\circ , \label{eq1210-01}\\
|V^0| &= 
\left(
\begin{array}{ccc}
0.9741-0.9749 & 0.2226-0.2259 & 0.0024-0.0038 \\
0.2225-0.2259 & 0.9734-0.9745 & 0.0295-0.0387 \\
0.0048-0.0084 & 0.0289-0.0379 & 0.9993-0.9996
\end{array}
\right). \label{eq1118-01} 
\end{align}
The ratios among CKM matrix elements are 
\begin{align}
|V_{cd}^0/V_{cs}^0| &= 0.2284-0.2320, & 
|V_{us}^0/V_{cs}^0| &= 0.2285-0.2321, \nonumber \\
|V_{td}^0/V_{ts}^0| &= 0.1699-0.2252, & 
|V_{td}^0/V_{tb}^0| &= 0.0050-0.0084, \nonumber \\
|V_{ub}^0/V_{cb}^0| &= 0.0747-0.1055, & 
|V_{ub}^0/V_{tb}^0| &= 0.0024-0.0038. \label{vratio}
\end{align}
These values are useful to derive the allowed regions in the parameters $\alpha_u$ and $\alpha_d$
as shown in next section.

\section{CKM quark mixing matrix}
\par
In our model, \(M_{d}\) and \(M_u\) have the same zero texture with same or different assignments as follows:
\begin{eqnarray}
M_d  & =& \left(
\begin{array}{lll}
\ 0 & \ a_de^{-i\phi_d} & \ a_d \\
\ a_d^\prime e^{-i\phi_d} & \ b_de^{-2i\phi_d} & \ (1-\xi_d)b_d \\
\ a_d^\prime  & \ (1-\xi_d)b_d & \ b_d \\
\end{array}
\right) , \\
M_u  & =& \left(
\begin{array}{lll}
\ 0 & \ a_ue^{-i\phi_u} & \ a_u \\
\ a_u^\prime e^{-i\phi_u} & \ b_ue^{-2i\phi_u} & \ (1-\xi_u)b_u \\
\ a_u^\prime  & \ (1-\xi_u)b_u & \ b_u \\
\end{array}
\right) , 
\end{eqnarray}
where \(\phi_{d}\) and \(\phi_u\) are phase parameters. 
We analyze the CKM quark mixing matrix 
of the model by taking the type A, the type B, and the type C assignments for up and down quarks.  
Now we present analytical results for only following case~(i), case~(ii), and case~(iii), since
the results for other cases are similarly obtained:
\par 
Case~(i): Type A assignment for up quarks and type B for down quarks are taken.  
\par 
Case~(ii): Type B assignment for up quarks and type A for down quarks are taken. 
\par
Case~(iii): Type A assignment both for up quarks and down quarks are taken. \\
\noindent As shown later, only case~(i) provides reasonable predictions for the CKM quark mixing matrix which are consistent with  Eq.~(\ref{eq1210-01}). 
For other cases we have no consistent predictions if we use the center values of running quark masses.
\par
In the following discussions, we denote the quark masses $(m_{1f},m_{2f},m_{3f})$ as $(m_u,m_c,m_t)$ for $f=u$, and 
as $(m_d,m_s,m_b)$ for $f=d$. 

\subsection{Type A for up $\times$ Type B for down}
\par
In case~(i), the CKM quark mixing matrix \(V\) is given by
\begin{eqnarray}
V&=&U^\dagger_{Lu}U_{Ld}=O^{T}_uP_{u}P^\dagger_{d} O^{\prime}_d=O_u^{T} P O^{\prime}_d\nonumber\\[.1in]
& =&
\left(
\begin{array}{ccc}
c_u c^\prime_d+\rho s_u s^\prime_d \quad & {\sigma}s_u 
\quad & c_u s^\prime_d-\rho s_u c^\prime_d \\
s_u c^\prime_d-\rho c_u s^\prime_d \quad & -{\sigma}c_u \quad & s_u s^\prime_d+{\rho}c_u c^\prime_d\\
-{\sigma}s^\prime_d \quad & -\rho  
\quad & {\sigma}c^\prime_d \\
\end{array}
\right)\equiv V^0,\label{ourckm-AB} 
\end{eqnarray}
where 
\begin{align}
s_u &= \sqrt{\frac{\frac{|m_{u}|m_{c}}{\alpha_u}-m_{u}^2}{m_{c}^2-m_{u}^2}}, &
c_u &= \sqrt{\frac{m_{c}^2-\frac{|m_{u}|m_{c}}{\alpha_u}}{m_{c}^2-m_{u}^2}},  
\nonumber\\
s^\prime_d &= \sqrt{\frac{\frac{|m_{d}|m_{b}}{\alpha_d}-m_{d}^2}{m_{b}^2-m_{d}^2}}, &
c^\prime_d &= \sqrt{\frac{m_{b}^2-\frac{|m_{d}|m_{b}}{\alpha_d}}{m_{b}^2-m_{d}^2}}. 
\end{align}
Here we have put 
\begin{equation}
P \equiv P_{u}P^\dagger_{d} \equiv \mbox{diag}(1, e^{i\delta_2},e^{i\delta_3}) , 
\label{P}
\end{equation}
and $O_u$ and $O^\prime_d$ are shown in Eq.~(\ref{eq990114}) 
and Eq.~(\ref{eq990114B}), respectively.
The \(\rho\) and \(\sigma\) are defined by 
\begin{eqnarray}
\rho & =&\frac{1}{2}(e^{i\delta_3}+e^{i\delta_2})
=\cos\frac{\delta_3 - \delta_2}{2} \exp i
\left( \frac{\delta_3 + \delta_2}{2} \right) \ ,\label{rho}\\ 
\sigma & =&\frac{1}{2}(e^{i\delta_3}-e^{i\delta_2})
= \sin\frac{\delta_3 - \delta_2}{2} 
\exp i \left( \frac{\delta_3 + \delta_2}{2}+ \frac{\pi}{2}
\right) \ . \label{sigma}
\end{eqnarray}
By using Eqs.~(\ref{P}) and (\ref{pf}), the phases $\delta_2$ and $\delta_3$ 
in our model are given by
\begin{eqnarray}
\delta_2 & =&\varphi_d-\varphi_u-(\phi_d-\phi_u), \label{delta2}\\
\delta_3 & =& \varphi_d-\varphi_u . \label{delta3}
\end{eqnarray}
Note also that we obtain phase-parameter-independent relations, 
\begin{eqnarray}
\frac{\left|V^0_{us}\right|}{\left|V^0_{cs}\right|} & =& \frac{s_u}{c_u}
	= \sqrt{\frac{|m_u|}{m_c}}\sqrt{\frac{\frac{m_c}{\alpha_u}-|m_u|}{m_c-\frac{|m_u|}{\alpha_u}}}, 
	\label{relationAB-1}\\
\frac{\left|V^0_{td}\right|}{\left|V^0_{tb}\right|} & =&  \frac{s^\prime_d}{c^\prime_d}
	= \sqrt{\frac{|m_d|}{m_b}}\sqrt{\frac{\frac{m_b}{\alpha_d}-|m_d|}{m_b-\frac{|m_d|}{\alpha_d}}}. 
	\label{relationAB-2}
\end{eqnarray}
Therefore, from Eqs.~(\ref{relationAB-1}) and (\ref{relationAB-2}) with Eq.~(\ref{vratio}), 
we can fix the parameters $\alpha_u$ and $\alpha_d$ 
as
\begin{eqnarray}
\alpha_u  &= &	\frac{\left|\frac{V^0_{us}}{V^0_{cs}}\right|^2+\frac{m_u^2}{m_c^2}}
		{\frac{|m_u|}{m_c}\left(1+\left|\frac{V^0_{us}}{V^0_{cs}}\right|^2\right)}
		=+0.0674 \sim +0.0694 \ (+0.0503 \sim +0.0901), \label{alpha-u-AB}\\
\alpha_d  &= & 	\frac{\left|\frac{V^0_{td}}{V^0_{tb}}\right|^2+\frac{m_d^2}{m_b^2}}
		{\frac{|m_d|}{m_b}\left(1+\left|\frac{V^0_{td}}{V^0_{tb}}\right|^2\right)}
		=+18.3303 \sim +50.1285 \ (+15.2264 \sim +57.4358). \label{alpha-d-AB}
\end{eqnarray}
Here, the values out of the parentheses are derived by 
using the center values of quark masses in Eq.~(\ref{eq123104}).
On the other hand, 
the values in the parentheses are obtained by using quark mass values 
in the ranges with the estimation errors in Eq.~(\ref{eq123104}).
\par
The phases $\phi_u$, $\varphi_u$, $\phi_d$, and $\varphi_d$ are given by 
\begin{eqnarray}
\cos \phi_u & =& \frac{|X_u|^2-m_{t}^2}
	{\sqrt{\left(|X_u|^2+m_{t}^2\right)^2-4m_{t}^2|X_u|^2\cos^2 \eta_u}} ,\label{phi-u-AB}\\
\cos \phi_d & =&  \frac{|X_d|^2-m_{s}^2}
	{\sqrt{\left(|X_d|^2+m_{s}^2\right)^2-4m_{s}^2|X_d|^2\cos^2 \eta_d}}  ,\label{phi-d-AB}\\
\cos \varphi_u & =& \frac{|X_u|-m_{t}\cos \eta_u}
	{\sqrt{|X_u|^2+m_{t}^2-2m_{t}|X_u|\cos \eta_u}} ,\\
\cos \varphi_d & =&  \frac{|X_d|-m_{s}\cos \eta_d}
	{\sqrt{|X_d|^2+m_{s}^2-2m_{s}|X_d|\cos \eta_d}} ,
\end{eqnarray}
where $|X_u|^2$ and $|X_d|^2$ are given by 
\begin{eqnarray}
|X_u|^2 & =&m_{u}^2+m_{c}^2-|m_{u}|m_{c}\left(\frac{1+\alpha_u^2}{\alpha_u}\right) ,\\
|X_d|^2 & =&m_{d}^2+m_{b}^2-|m_{d}|m_{b}\left(\frac{1+\alpha_d^2}{\alpha_d}\right) . 
\end{eqnarray}
Therefore, we obtain
\begin{eqnarray}
\phi_u &=& +3.1370 \sim +3.1462 \ (+3.1344 \sim +3.1488), \\
\phi_d &=& -0.0548 \sim +0.0548 \ (-0.0650 \sim +0.0650),
\end{eqnarray}
from Eqs.~(\ref{phi-u-AB}) -- (\ref{phi-d-AB}) with use of the restrictions in Eqs.~(\ref{alpha-u-AB})--(\ref{alpha-d-AB}) 
and free  $\eta_u$ and $\eta_d$ (i.e. $0 \le \eta_u \le 2\pi$ and $0 \le  \eta_d \le 2\pi$). 
It should be noted that all components of $V$ are dependent on 
$\delta_3$, $\delta_2$, $\alpha_u$, and $\alpha_d$. 
The phase parameters $\delta_3$ and $\delta_2$ are dependent on 
$\alpha_u$, $\alpha_d$, $\eta_u$, and $\eta_d$. 
Therefore, the present model has four adjustable parameters, 
$\alpha_u$, $\alpha_d$, $\eta_u$, and  $\eta_d$.
Then, the explicit magnitudes of the components of \(V\) are obtained as 
\begin{eqnarray}
\left|V^0_{cs}\right|& =&\left|\sigma c_u\right| 
	= \left|c_u \sin \left( \frac{\delta_3-\delta_2}{2}\right) \right|
	= 0.9737-0.9749 \ (0.9539-0.9852), \label{eq3024-AB}\\
\left| V^0_{us}\right|& =&\left|\sigma s_u\right| 
	= \left|s_u \sin \left( \frac{\delta_3-\delta_2}{2}\right) \right|
	= 0.2226-0.2260  \ (0.1687-0.2981), \label{eq3025}\\
\left| V^0_{td}\right|& =&\left|\sigma s^\prime_d \right|
	= \left|s^\prime_d \sin \left(\frac{\delta_3-\delta_2}{2}\right) \right|
	= 0.0050-0.0084  \ (0.0042-0.0100), \label{eq3026}\\
\left|V^0_{ts}\right|& =&\left|\rho \right| 
	= \left|\cos \left(\frac{\delta_3 - \delta_2}{2}\right)\right|
	= 0.0000 - 0.0297 \ (0.0000 - 0.0361), \label{eq3027}\\
\left|V^0_{tb}\right|& =&\left|\sigma c^\prime_d\right|
	= \left|c^\prime_d \sin \left(\frac{\delta_3-\delta_2}{2}\right)\right| 
	= 0.9995-1.0000  \ (0.9993-1.0000). \label{eq3028-AB}
\end{eqnarray}
By using these values out of the parentheses 
which are obtained by using the center values of quark masses in Eq.~(\ref{eq123104}), 
the quark mixing angles are predicted as follows:
\begin{align}
\sin \theta_{12}^0 &= 0.2226-0.2260, & 
\sin \theta_{23}^0 &= 0.0000-0.0303, &
\sin \theta_{13}^0 &= 0.0000-0.0137.
\end{align}
These predicted values are consistent with Eq.~(\ref{eq1210-01}).
\par
By using the rephasing of the up and down quarks, 
Eq.~(\ref{ourckm-AB}) is changed to the standard representation of the CKM quark mixing matrix, 
\begin{eqnarray}
V_{\rm std} &=& \mbox{diag}(e^{i\zeta_1^u},e^{i\zeta_2^u},e^{i\zeta_2^u})  \ V \ 
\mbox{diag}(e^{i\zeta_1^d},e^{i\zeta_2^d},e^{i\zeta_2^d}) \nonumber \\
&=&
\left(
\begin{array}{ccc}
c_{13}c_{12} & c_{13}s_{12} & s_{13}e^{-i\delta} \\
-c_{23}s_{12}-s_{23}c_{12}s_{13} e^{i\delta}
&c_{23}c_{12}-s_{23}s_{12}s_{13} e^{i\delta} 
&s_{23}c_{13} \\
s_{23}s_{12}-c_{23}c_{12}s_{13} e^{i\delta}
 & -s_{23}c_{12}-c_{23}s_{12}s_{13} e^{i\delta} 
& c_{23}c_{13} \\
\end{array}
\right) \ .
\label{stdrep}
\end{eqnarray}
Here \(\zeta_i^q\) comes from the rephasing in the quark fields 
to make the choice of phase convention.
The $CP$ violating phase \(\delta^0\) in Eq.~(\ref{stdrep}) 
is predicted with the expression of $V^0$ in Eq.~(\ref{ourckm-AB}) as 
\begin{equation}
\delta^0 =
\mbox{arg}\left[
\left(\frac{V_{us}^0 V_{cs}^{0*}}{V_{ub}^0 V_{cb}^{0*}}\right) + 
\frac{|V_{us}^0|^2}{1-|V_{ub}^0|^2}
\right]= 0 - 2\pi\ .
\end{equation}
\par
So far we have predicted magnitudes for the  components of $|V_{ij}^0|$ and \(\delta^0\) 
by using Eqs.~(\ref{alpha-u-AB})--(\ref{alpha-d-AB}) 
and free parameters $\eta_u$ and $\eta_d$ (i.e. $0 \le \eta_u \le 2\pi$ and $0 \le  \eta_d \le 2\pi$). 
Conversely, let us derive the allowed regions of the parameters, \(\alpha_u\), \(\alpha_d\), \(\eta_u\), and \(\eta_d\) from the experimental restriction of Eqs.~(\ref{eq1210-01}) - (\ref{vratio}).  
In Fig.~2, we present the allowed regions of the parameters, \(\alpha_u\), \(\alpha_d\), \(\eta_u\), and \(\eta_d\) in which the experimental restrictions of Eqs.~(\ref{eq1210-01}) - (\ref{vratio}) are satisfied simultaneously. 
In Fig.~3, with use of free $\eta_u$ and $\eta_d$, we also obtain the allowed regions in the $\alpha_u-\alpha_d$ parameter space which 
come from the restrictions of $|V^0_{us}|$, $|V^0_{cb}|$, $|V^0_{ub}|$,  
and the $CP$ violating phase \(\delta^0\) in Eqs.~(\ref{eq1210-01}) - (\ref{eq1118-01}), respectively. 
As seen from Fig.~2 and Fig.~3, the case~(i) is well consistent with the observed experimental data.

\subsection{Type B for up $\times$ Type A for down}
In case~(ii), the CKM quark mixing matrix \(V\) is given by
\begin{eqnarray}
V&=&U^\dagger_{Lu}U_{Ld}=O^{\prime T}_uP_{u}P^\dagger_{d} O_d=O_u^{\prime T} P O_d\nonumber\\[.1in]
& =&
\left(
\begin{array}{ccc}
c^\prime_uc_d+\rho s^\prime_u s_d \quad & c^\prime_u s_d-\rho s^\prime_u c_d 
\quad & -{\sigma}s^\prime_u \\
{\sigma}s_d \quad & -{\sigma}c_d \quad & -\rho \\
s^\prime_u c_d-{\rho}c^\prime_u s_d \quad & s^\prime_u s_d+{\rho}c^\prime_u c_d 
\quad & {\sigma}c^\prime_u \\
\end{array}
\right)\equiv V^0,\label{ourckm-BA} 
\end{eqnarray}
where 
\begin{align}
s^\prime_u &= \sqrt{\frac{\frac{|m_{u}|m_{t}}
	{\alpha_u}-m_{u}^2}{m_{t}^2-m_{u}^2}},&
c^\prime_u &= \sqrt{\frac{m_{t}^2-\frac{|m_{u}|m_{t}}{\alpha_u}}
	{m_{t}^2-m_{u}^2}},  \nonumber \\
s_d &= \sqrt{\frac{\frac{|m_{d}|m_{s}}{\alpha_d}-m_{d}^2}
	{m_{s}^2-m_{d}^2}},&
c_d &= \sqrt{\frac{m_{s}^2-\frac{|m_{d}|m_{s}}{\alpha_d}}
	{m_{s}^2-m_{d}^2}}, 
\end{align}
and $O_d$ and $O^\prime_u$ are shown in Eq.~(\ref{eq990114}) and 
Eq.~(\ref{eq990114B}), respectively.
Using the phase-parameter-independent relations, 
\begin{eqnarray}
\frac{\left|V^0_{ub}\right|}{\left|V^0_{tb}\right|}  & =& \frac{s^\prime_u}{c^\prime_u} = \sqrt{\frac{|m_u|}{m_t}}\sqrt{\frac{\frac{m_t}{\alpha_u}-|m_u|}{m_t-\frac{|m_u|}{\alpha_u}}},\label{relationBA-1}\\
\frac{\left|V^0_{cd}\right|}{\left|V^0_{cs}\right|}  & =& \frac{s_d}{c_d} = \sqrt{\frac{|m_d|}{m_s}}\sqrt{\frac{\frac{m_s}{\alpha_d}-|m_d|}{m_s-\frac{|m_d|}{\alpha_d}}},\label{relationBA-2}
\end{eqnarray}
with Eq.~(\ref{vratio}), we obtain
\begin{eqnarray}
\alpha_u  &= & 	\frac{\left|\frac{V^0_{ub}}{V^0_{tb}}\right|^2+\frac{m_u^2}{m_t^2}}
		{\frac{|m_u|}{m_t}\left(1+\left|\frac{V^0_{ub}}{V^0_{tb}}\right|^2\right)}
		=+0.5706 \sim +1.4294 \ (+0.1829 \sim +2.4503), \label{alpha-u-BA}\\
\alpha_d  &= &	\frac{\left|\frac{V^0_{cd}}{V^0_{cs}}\right|^2+\frac{m_d^2}{m_s^2}}
		{\frac{|m_d|}{m_s}\left(1+\left|\frac{V^0_{cd}}{V^0_{cs}}\right|^2\right)}
		=+0.9388 \sim +0.9658 \ (+0.7293 \sim +1.2255). \label{alpha-d-BA}
\end{eqnarray}
\par
The phases $\phi_u$, $\varphi_u$, $\phi_d$, and $\varphi_d$ are given by 
\begin{eqnarray}
\cos \phi_u & =& \frac{|X_u|^2-m_{c}^2}
	{\sqrt{\left(|X_u|^2+m_{c}^2\right)^2-4m_{c}^2|X_u|^2\cos^2 \eta_u}} ,\label{phi-u-BA}\\
\cos \phi_d & =&  \frac{|X_d|^2-m_{b}^2}
	{\sqrt{\left(|X_d|^2+m_{b}^2\right)^2-4m_{b}^2|X_d|^2\cos^2 \eta_d}}  ,\label{phi-d-BA}\\
\cos \varphi_u & =& \frac{|X_u|-m_{c}\cos \eta_u}
	{\sqrt{|X_u|^2+m_{c}^2-2m_{c}|X_u|\cos \eta_u}} ,\\
\cos \varphi_d & =&  \frac{|X_d|-m_{b}\cos \eta_d}
	{\sqrt{|X_d|^2+m_{b}^2-2m_{b}|X_d|\cos \eta_d}} ,
\end{eqnarray}
where 
\begin{eqnarray}
|X_u|^2 & =&m_{u}^2+m_{t}^2-|m_{u}|m_{t}\left(\frac{1+\alpha_u^2}{\alpha_u}\right) ,\\
|X_d|^2 & =&m_{d}^2+m_{s}^2-|m_{d}|m_{s}\left(\frac{1+\alpha_d^2}{\alpha_d}\right) 
\label{X-d-BA}. 
\end{eqnarray}
Therefore, we obtain
\begin{eqnarray}
\phi_u &=& -0.0047 \sim +0.0047 \ (-0.0073 \sim +0.0073), \\
\phi_d &=& +3.0913 \sim +3.1919 \ (+3.0819 \sim +3.2013), 
\end{eqnarray}
from Eqs.~(\ref{phi-u-BA})--(\ref{phi-d-BA}) with use of the restrictions of $\alpha_u$ and $\alpha_d$ 
in Eqs.~(\ref{alpha-u-BA})--(\ref{alpha-d-BA}) and free $\eta_u$ and $\eta_d$ (i.e. $0 \le \eta_u \le 2\pi$ and $0 \le  \eta_d \le 2\pi$). 
Hence, the explicit magnitudes of the components of \(V\) are obtained as 
\begin{eqnarray}
\left|V^0_{cb}\right|& =&\left|\rho\right|
	= \left|\cos\left(\frac{\delta_3-\delta_2}{2}\right)\right|
	= 0.0000 - 0.0275 \ (0.0000 - 0.0335) ,\label{eq3021}\\
\left|V^0_{ub}\right|& =&\left|\sigma s^\prime_u \right| 
	= \left|\sin\left(\frac{\delta_3-\delta_2}{2}\right) s^\prime_u\right|
	= 0.0024 - 0.0038 \ (0.0010 - 0.0087) ,\\
\left|V^0_{cd}\right|& =&\left|\sigma s_d \right| 
	= \left|\sin\left(\frac{\delta_3-\delta_2}{2}\right) s_d\right|
	= 0.2226 - 0.2260 \ (0.1725 - 0.2937) ,\label{eq3023}\\
\left|V^0_{cs}\right|& =&\left|\sigma c_d \right|
	= \left|\sin\left(\frac{\delta_3-\delta_2}{2}\right) c_d\right|
	= 0.9738 - 0.9749 \ (0.9556 - 0.9850), \label{eq3024-BA}\\
\left|V^0_{tb}\right|& =&\left|\sigma c^\prime_u \right|
	= \left|\sin\left(\frac{\delta_3-\delta_2}{2}\right) c^\prime_u\right|
	= 0.9996 - 0.9997 \ (0.9994 - 0.9997)  . \label{eq3028-BA}
\end{eqnarray}
Here the values in parentheses, 
which are obtained by using the quark masses in the ranges with the estimation errors in Eq.~(\ref{eq123104}), 
are consistent with Eq.~(\ref{eq1118-01}).
However, the values out of parentheses are not. 
Namely, when we use the center values of quark masses in Eq.~(\ref{eq123104}), 
the quark mixing angles are predicted as
\begin{align}
\sin\theta_{12}^0 &= 0.2225-0.2260, &
\sin\theta_{23}^0 &= 0.0000-0.0275, \nonumber \\
\sin\theta_{13}^0 &= 0.0024-0.0038, &
\delta^0  \      &= -3.0896 \ - \  + 3.0841 ,
\end{align}
which are inconsistent with Eq.~(\ref{eq1210-01}).
In Fig.~4, allowed regions in the $\alpha_u-\alpha_d$ parameter space are derived, which 
are obtained from the restrictions of $|V^0_{us}|$, $|V^0_{cb}|$, $|V^0_{ub}|$,  
and the $CP$ violating phase \(\delta^0\) in Eqs.~(\ref{eq1210-01}) - (\ref{eq1118-01}), respectively 
with use of free $\eta_u$ and $\eta_d$ (i.e. $0 \le \eta_u \le 2\pi$ and $0 \le  \eta_d \le 2\pi$). 
As seen from Fig.~4, the case (ii) is inconsistent with Eq.~(\ref{eq1118-01}) 
if we use the center values of the quark masses.

\par
\subsection{Type A for up $\times$ Type A for down}
\par 
In case (iii), the CKM quark mixing matrix \(V\) is given by
\begin{eqnarray}
V&=&U^\dagger_{Lu}U_{Ld}=O^T_uP_{u}P^\dagger_{d} O_d\nonumber\\[.1in]
& =&
\left(
\begin{array}{ccc}
c_uc_d+\rho s_u s_d \quad & c_u s_d-\rho s_u c_d \quad & -{\sigma}s_u \\
s_uc_d-\rho c_u s_d \quad & s_u s_d+\rho c_u c_d \quad & {\sigma}c_u \\
-{\sigma}s_d \quad & {\sigma}c_d \quad & \rho \\
\end{array}
\right)\equiv V^0,\label{ourckm-AA} 
\end{eqnarray}
where 
\begin{align}
s_u &= \sqrt{\frac{\frac{|m_{u}|m_{c}}
	{\alpha_u}-m_{u}^2}{m_{c}^2-m_{u}^2}},&
c_u &= \sqrt{\frac{m_{c}^2-\frac{|m_{u}|m_{c}}
	{\alpha_u}}{m_{c}^2-m_{u}^2}}, \nonumber \\
s_d &= \sqrt{\frac{\frac{|m_{d}|m_{s}}
	{\alpha_d}-m_{d}^2}{m_{s}^2-m_{d}^2}},&
c_d &= \sqrt{\frac{m_{s}^2-\frac{|m_{d}|m_{s}}{\alpha_d}}
	{m_{s}^2-m_{d}^2}}, 
\end{align}
and $O_d$ ($O_u$) is shown in Eq.~(\ref{eq990114}). 
Here we have put \(P \equiv P_{u}P^\dagger_{d} \equiv 
\mbox{diag}(1, e^{i\delta_2},e^{i\delta_3})\). 
The \(\rho\) and \(\sigma\) are defined by Eqs.~(\ref{rho}) and (\ref{sigma}), 
respectively.
Using the phase-parameter-independent relations, 
\begin{eqnarray}
\frac{|V^0_{ub}|}{|V^0_{cb}|} & =& \frac{s_u}{c_u}
	= \sqrt{\frac{|m_u|}{m_c}}\sqrt{\frac{\frac{m_c}{\alpha_u}-|m_u|}
		{m_c-\frac{|m_u|}{\alpha_u}}} , \label{relationAA-1}\\
\frac{|V^0_{td}|}{|V^0_{ts}|} & =&  \frac{s_d}{c_d}
	= \sqrt{\frac{|m_d|}{m_s}}\sqrt{\frac{\frac{m_s}{\alpha_d}-|m_d|}
		{m_s-\frac{|m_d|}{\alpha_d}}} , \label{relationAA-2}
\end{eqnarray} 
with Eq.~(\ref{vratio}), we obtain
\begin{eqnarray}
\alpha_u  &= &  \frac{\left|\frac{V^0_{ub}}{V^0_{cb}}\right|^2+\frac{m_u^2}{m_c^2}}
		{\frac{|m_u|}{m_c}\left(1+\left|\frac{V^0_{ub}}{V^0_{cb}}\right|^2\right)}
		= +0.3123 \sim +0.6188 \ (+0.2330 \sim +0.8026), \label{alpha-u-AA}\\
\alpha_d  &= &  \frac{\left|\frac{V^0_{td}}{V^0_{ts}}\right|^2+\frac{m_d^2}{m_s^2}}
		{\frac{|m_d|}{m_s}\left(1+\left|\frac{V^0_{td}}{V^0_{ts}}\right|^2\right)}
		= +0.9907 \sim +1.6458 \ (+0.7704 \sim +2.0398). \label{alpha-d-AA}
\end{eqnarray}
The phases $\phi_u$, $\varphi_u$, $\phi_d$, and $\varphi_d$ are given by 
\begin{eqnarray}
\cos \phi_u & =& \frac{|X_u|^2-m_{t}^2}
	{\sqrt{\left(|X_u|^2+m_{t}^2\right)^2-4m_{t}^2|X_u|^2\cos^2 \eta_u}} ,\label{phi-u-AA}\\ 
\cos \phi_d & =&  \frac{|X_d|^2-m_{b}^2}
	{\sqrt{\left(|X_d|^2+m_{b}^2\right)^2-4m_{b}^2|X_d|^2\cos^2 \eta_d}}  ,\label{phi-d-AA}\\
\cos \varphi_u & =& \frac{|X_u|-m_{t}\cos \eta_u}
	{\sqrt{|X_u|^2+m_{t}^2-2m_{t}|X_u|\cos \eta_u}} ,\\
\cos \varphi_d & =&  \frac{|X_d|-m_{b}\cos \eta_d}
	{\sqrt{|X_d|^2+m_{b}^2-2m_{b}|X_d|\cos \eta_d}} ,
\end{eqnarray}
where 
\begin{eqnarray}
|X_u|^2 & =&m_{u}^2+m_{c}^2-|m_{u}|m_{c}\left(\frac{1+\alpha_u^2}{\alpha_u}\right) , \\
|X_d|^2 & =&m_{d}^2+m_{s}^2-|m_{d}|m_{s}\left(\frac{1+\alpha_d^2}{\alpha_d}\right) .
\label{X_d_AA} 
\end{eqnarray}
Therefore, we obtain 
\begin{eqnarray}
\phi_u &=& +3.1369 \sim +3.1463 \ (+3.1343 \sim +3.1489), \\
\phi_d &=& +3.0913 \sim +3.1919 \ (+3.0819 \sim +3.2013),
\end{eqnarray}
from Eqs.~(\ref{phi-u-AA})--(\ref{phi-d-AA}) with use of the restrictions of $\alpha_u$ and $\alpha_d$ 
in Eqs.~(\ref{alpha-u-AA})--(\ref{alpha-d-AA}) and free $\eta_u$ and $\eta_d$ (i.e. $0 \le \eta_u \le 2\pi$ and $0 \le  \eta_d \le 2\pi$). 
Hence, the explicit magnitudes of the components of \(V\) are obtained as 
\begin{eqnarray}
\left|V^0_{cb} \right|& =&\left|\sigma c_u\right|
	= \left|c_u \sin \left( \frac{\delta_3-\delta_2}{2}\right)\right|
	= 0.0000-0.0274 \ (0.0000-0.0334),\label{eq5021}\\
\left|V^0_{ub}\right| & =&\left|\sigma s_u \right| 
	= \left|s_u \sin \left( \frac{\delta_3-\delta_2}{2}\right)\right|
	= 0.0000-0.0029 \ (0.0000-0.0046), \label{eq5022}\\
\left| V^0_{td}\right| & =&\left|\sigma s_d \right|
	= \left|s_d \sin \left(\frac{\delta_3 - \delta_2}{2}\right) \right|
	= 0.0000-0.0060 \ (0.0000-0.0083),\label{eq5026}\\
\left|V^0_{ts}\right| & =&\left|\sigma c_d \right|
	= \left|c_d \sin \left(\frac{\delta_3 - \delta_2}{2} \right) \right|
	= 0.0000-0.0269 \ (0.0000-0.0329),\label{eq5027}\\
\left|V^0_{tb}\right| & =&\left|\rho\right| 
	= \left|\cos \left(\frac{\delta_3-\delta_2}{2} \right)\right|
	= 0.9996 - 1.0000 \ (0.9994 - 1.0000).\label{eq5028}
\end{eqnarray}
Here the values in parentheses, 
which are obtained by using the quark mass values in the ranges with the estimation errors in Eq.~(\ref{eq123104}), 
are consistent with Eq.~(\ref{eq1118-01}).
However, the values out of parentheses are not. 
Namely, when we use the center values of quark masses in Eq.~(\ref{eq123104}), 
the quark mixing angles are predicted as  
\begin{align}
\sin\theta_{12}^0 &= 0.0631-0.3208, &
\sin\theta_{23}^0 &= 0.0000-0.0274, \nonumber \\
\sin\theta_{13}^0 &= 0.0000-0.0029, &
\delta^0 \      &= 0 - 2\pi, 
\end{align}
which are inconsistent with Eq.~(\ref{eq1210-01}).
In Fig.~5, allowed regions in the $\alpha_u-\alpha_d$ parameter space are derived, which 
are obtained from the restrictions of $|V^0_{us}|$, $|V^0_{cb}|$, $|V^0_{ub}|$,  
and the $CP$ violating phase \(\delta^0\) in Eqs.~(\ref{eq1210-01}) - (\ref{eq1118-01}), respectively 
with use of free $\eta_u$ and $\eta_d$ (i.e. $0 \le \eta_u \le 2\pi$ and $0 \le  \eta_d \le 2\pi$). 
As seen from Fig.~5, the case (iii) is inconsistent with Eq.~(\ref{eq1118-01}) 
if we use the center values of the quark masses.


We have presented the results for only the case~(i), case~(ii), and case~(iii).  
For other possible cases we have no consistent allowed regions of the parameters.

\section{conclusion}
We have proposed a mass matrix model that gives a universal description of 
quark and lepton with the same texture form (\ref{ourMf}), 
based on a flavor $2 \leftrightarrow 3$ symmetry including a phase $\phi$.
Against the conventional picture that the mass matrix forms in the quark sector will 
take somewhat different structures from those in the lepton sector, 
we have considered a scenario in which the flavor $2 \leftrightarrow 3$ symmetry play a role  
in the quark mass matrices as well as in the lepton mass matrices 
as a fundamental common symmetry in the Yukawa couplings. 
By using the quark masses as inputs, the present model has four adjustable parameters, 
$\alpha_u$, $\alpha_d$, $\eta_u$, and  $\eta_d$ 
to reproduce the observed CKM quark mixing matrix parameters at the electroweak scale. 
We have shown that the observed small CKM quark mixings are realized by 
a fine tuning of the parameters in the case (i) where type A assignment for up quarks and type B for down quarks are taken. 
With this consistent parameters, our result suggests an interesting mass matrix structure for up and down quarks, 
which will give a clue to a possible form of mass matrix.  
Since the components of each mass matrix $M_f$ takes different values, 
the present model cannot be embedded into a GUT scenario. 
However, it is worth-while noting that it can give a universal description of 
quark and lepton mass matrices with the same texture form 
based on a flavor $2 \leftrightarrow 3$ symmetry.



\begin{acknowledgments}
We would like to thank Y. Koide for helpful discussions. This work of K.M. was supported by the JSPS, No. 3700.
\end{acknowledgments}
\appendix*

\section{}
Following the discussions in Ref.~\cite{Matsuda}, 
let us give a brief review of the diagonalization of mass matrix $\widehat{M}$ given by 
\begin{equation}
\widehat{M}=\left(
\begin{array}{lll}
\ 0 & \ A_f & \ A_f \\
\ A_f & \ B_f & \ C_f \\
\ A_f & \ C_f & \ B_f \\
\end{array}
\right).
\end{equation}
The eigenvalues of $\widehat{M}$ are given by 
\begin{eqnarray}
-\lambda_{1f}& =&
\frac{1}{2}
\left[B_f+C_f-\sqrt{8A_f^2 + (B_f+C_f)^2}
\right] ,\\
\lambda_{2f}& =&\frac{1}{2}
\left[B_f+C_f+\sqrt{8A_f^2 + (B_f+C_f)^2}
\right] ,\\
\lambda_{3f}& =&B_f-C_f. 
\end{eqnarray}
To the contrary, the texture's components $A_f$, $B_f$, and $C_f$ are 
expressed in terms of $\lambda_{if}$ as 
\begin{eqnarray}
A_f & =& \sqrt{\frac{\lambda_{2f}\lambda_{1f}}{2}}  ,\nonumber\\
B_f & =&\frac{1}{2} \left(\lambda_{3f}+\lambda_{2f}-\lambda_{1f}\right) ,\label{eq2003}\\
C_f & =&-\frac{1}{2}\left(\lambda_{3f}-\lambda_{2f}+\lambda_{1f}\right) .\nonumber 
\end{eqnarray}
Here, there are following three types of assignments for the eigenvalues $\lambda_{if}$:
\par
(i) Type A: 
\par
In this type, the mass eigenvalues \(\lambda_{1f}\), \(\lambda_{2f}\), and \(\lambda_{3f}\) 
are allocated to the masses of the first, second, and third generations, respectively.
(i.e. \(\lambda_{1f}\)\(<\)\(\lambda_{2f}\)\(<\)\(\lambda_{3f}\).)
Therefore, $\widehat{M}$ is diagonalized by an orthogonal matrix $O_f$ as
\begin{equation}
O_f^T
\widehat{M}
O_f=\mbox{diag}(-\lambda_{1f},\lambda_{2f},\lambda_{3f}) ,
\end{equation}
with
\begin{equation}
O_f\equiv
\left(
\begin{array}{ccc}
{  c_f}  & {  s_f}& {0} \\
{- \frac{s_f}{\sqrt{2}}} & {\frac{c_f}{\sqrt{2}}} & {-\frac{1}{\sqrt{2}}} \\
{- \frac{s_f}{\sqrt{2}}} & {\frac{c_f}{\sqrt{2}}} & {\frac{1}{\sqrt{2}}}
\end{array}
\right),  
\end{equation}
where 
\begin{equation}
c_f  =\sqrt{\frac{\lambda_{2f}}{\lambda_{2f}+\lambda_{1f}}} ,\quad
s_f  =\sqrt{\frac{\lambda_{1f}}{\lambda_{2f}+\lambda_{1f}}}.
\end{equation}
\par
(ii) Type B: 
\par
In this type, the mass eigenvalues \(\lambda_{1f}\), \(\lambda_{3f}\), and \(\lambda_{2f}\) 
are allocated to the masses of the first, second, and third generations, respectively.
(i.e. \(\lambda_{1f}\)\(<\)\(\lambda_{3f}\)\(<\)\(\lambda_{2f}\).)
Therefore, $\widehat{M}$ is diagonalized by an orthogonal matrix $O_f^{\prime}$ as
\begin{equation}
O_f^{\prime T}
\widehat{M}
O_f^{\prime}=\mbox{diag}(-\lambda_{1f},\lambda_{3f},\lambda_{2f}) ,
\end{equation}
with
\begin{equation}
O_f^\prime\equiv
\left(
\begin{array}{ccc}
{ c_f} & {0} & { s_f} \\
{- \frac{s_f}{\sqrt{2}}} & {\frac{1}{\sqrt{2}}} & {\frac{c_f}{\sqrt{2}}} \\
{- \frac{s_f}{\sqrt{2}}} & {-\frac{1}{\sqrt{2}}} & {\frac{c_f}{\sqrt{2}}}
\end{array}
\right).  
\end{equation}
Here \(O_f^{\prime}\) is obtained from \(O_f\) by exchanging the second row for the third one. 
\par
(iii) Type C: 
\par
In this type, the mass eigenvalues \(\lambda_{3f}\), \(\lambda_{1f}\), and \(\lambda_{2f}\) 
are allocated to the masses of the first, second, and third generations, respectively.
(i.e. \(\lambda_{3f}\)\(<\)\(\lambda_{1f}\)\(<\)\(\lambda_{2f}\).)
Therefore, $\widehat{M}$ is diagonalized by an orthogonal matrix $O_f^{\prime \prime}$ as
\begin{equation}
O_f^{\prime \prime T}
\widehat{M}
O_f^{\prime \prime}=\mbox{diag}(\lambda_{3f},-\lambda_{1f},\lambda_{2f}) ,
\end{equation}
with
\begin{equation}
O_f^{\prime \prime}\equiv
\left(
\begin{array}{ccc}
{0}                   & { c_f}                   &  { s_f} \\
{\frac{1}{\sqrt{2}}}  & {- \frac{s_f}{\sqrt{2}}} &  {\frac{c_f}{\sqrt{2}}} \\
{-\frac{1}{\sqrt{2}}} & {- \frac{s_f}{\sqrt{2}}} &  {\frac{c_f}{\sqrt{2}}}
\end{array}
\right). 
\end{equation}
Here \(O_f^{\prime \prime}\) is obtained from \(O_f^{\prime}\) by exchanging the first row for the second one. 


\begin{figure}[htbp]
\begin{center}
\includegraphics{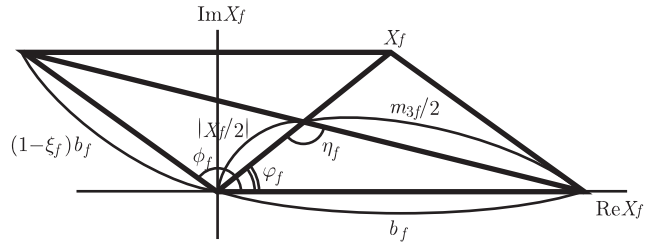}
\end{center}
\caption{%
The relations among the components in the mass matrix, and
the definition of a free $CP$ violating phase parameter $\eta_f$
which is independent of the mass eigenvalues.} 
\label{fig1}
\end{figure}

\begin{figure}[htbp]
\begin{center}
\includegraphics{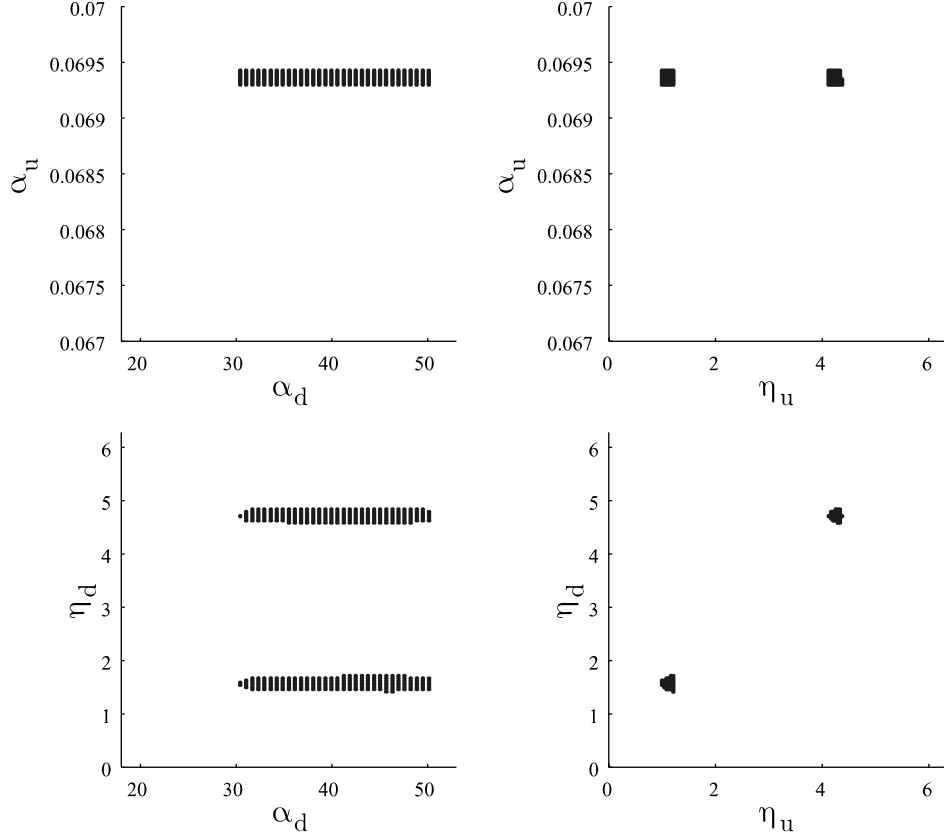}
\end{center}
\caption{%
The allowed region in the parameter planes 
for the case~(i) in which Type A assignment for up quarks and type B for 
down quarks are taken.
The dots indicate the allowed regions, 
in which the CKM quark mixing angles $\sin\theta_{12}^0$, 
$\sin\theta_{23}^0$, $\sin\theta_{13}^0$,  
and the $CP$ violating phase \(\delta^0\) at the GUT scale satisfy the restrictions of 
Eq.~(\ref{eq1210-01}), simultaneously.}
\label{fig2AB}
\end{figure}

\begin{figure}[htbp]
\begin{center}
\includegraphics{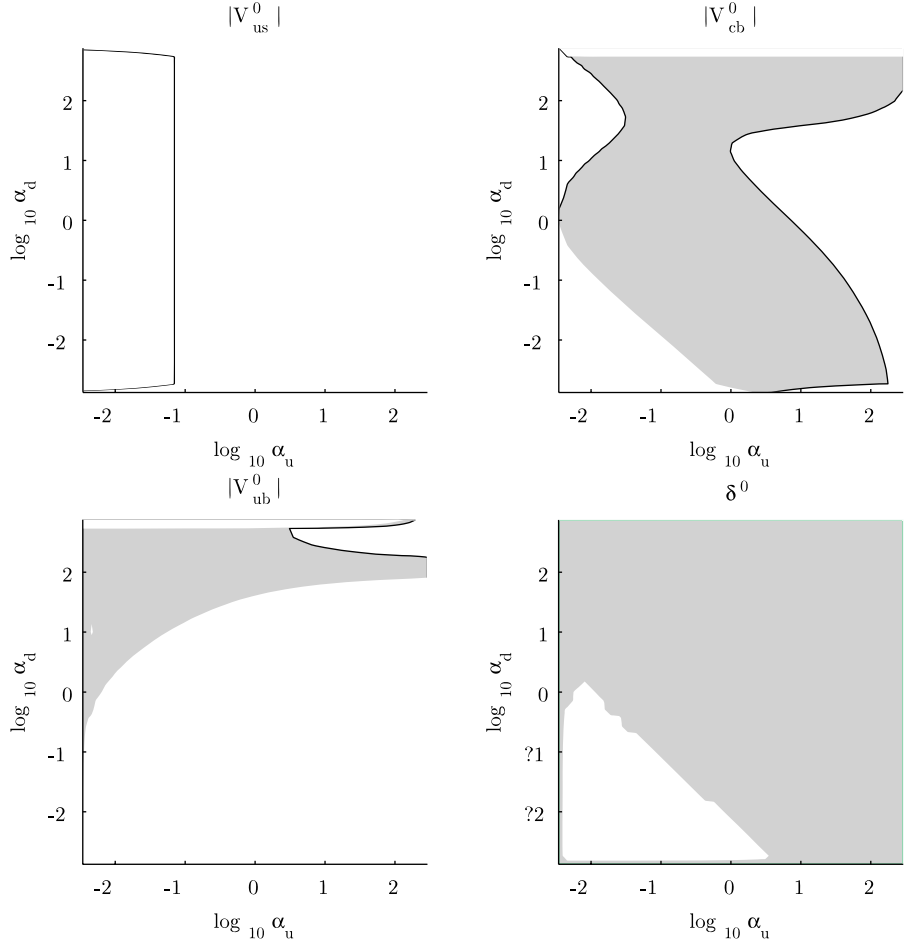}
\end{center}
\caption{%
The allowed region in the $\alpha_u$ - $\alpha_d$ parameter plane 
for the case~(i) in which Type A assignment for up quarks and type B for down quarks are taken.
The allowed regions are indicated by the dark shaded regions 
which are obtained from the values of the CKM quark mixing matrix elements $|V^0_{us}|$, $|V^0_{cb}|$, $|V^0_{ub}|$,  
and the $CP$ violating phase \(\delta^0\) in Eqs.~(\ref{eq1210-01}) - (\ref{eq1118-01}), respectively. 
The boundary with (without) black lines shows the upper (lower) limit. 
Here we take $\eta_u$ and $\eta_d$ as free parameters (i.e. $0 \le \eta_u \le 2\pi$ and $0 \le  \eta_d \le 2\pi$). }
\label{fig3AB}
\end{figure}
\begin{figure}[htbp]
\begin{center}
\includegraphics{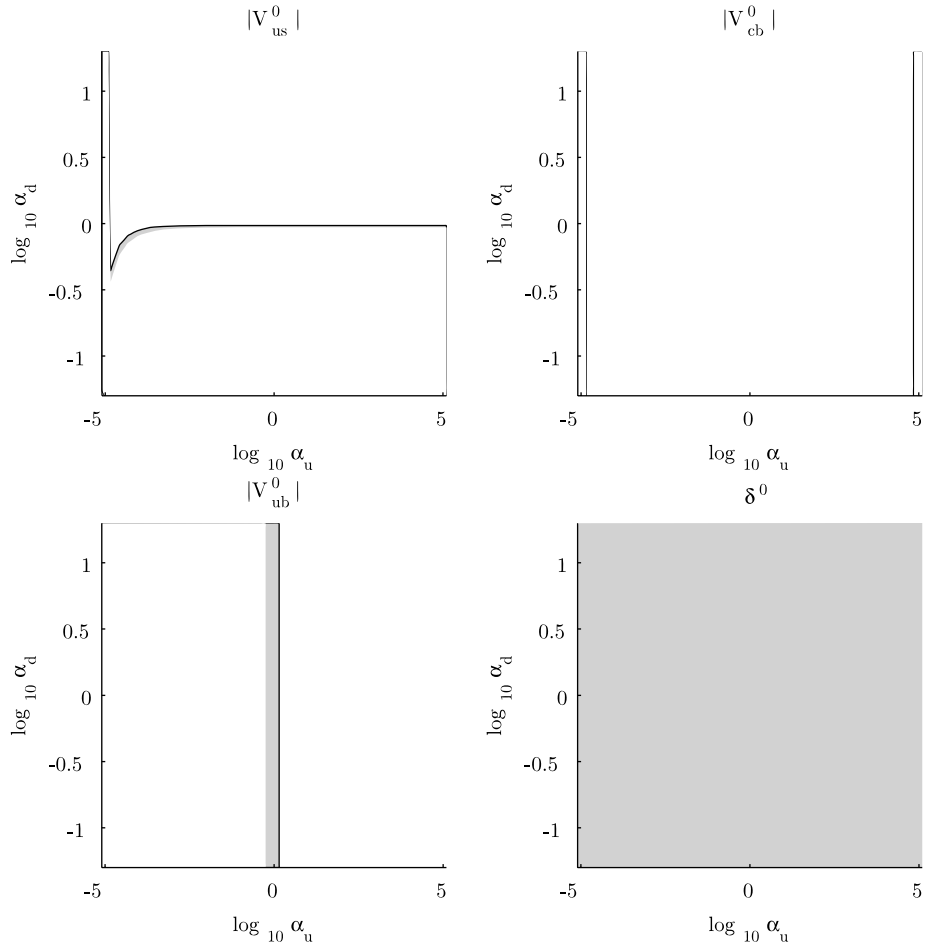}
\end{center}
\caption{%
The allowed region in the $\alpha_u$ - $\alpha_d$ parameter plane 
for the case~(ii) in which Type B assignment for up quarks and type A for down quarks are taken. 
}
\label{fig3BA}
\end{figure}
\begin{figure}[htbp]
\begin{center}
\includegraphics{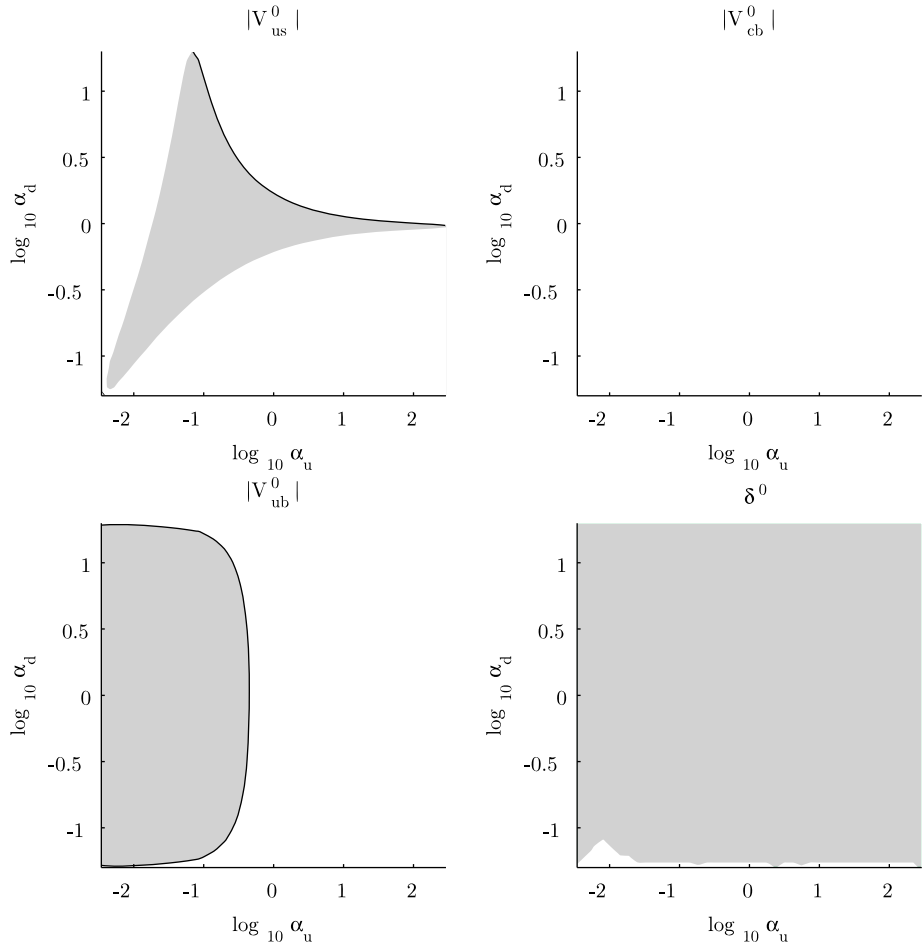}
\end{center}
\caption{%
The allowed region in the $\alpha_u$ - $\alpha_d$ parameter plane 
for the case~(iii) in which Type A assignment both for up quarks and down quarks are taken. 
}
\label{fig3AA}
\end{figure}

\end{document}